\documentclass[reprint,
nobibnotes,
 amsmath,amssymb,
 aps,prx
]{revtex4-2}

\usepackage[normalem]{ulem}
\usepackage{graphicx}%
\usepackage{subcaption}%
\usepackage{multirow}%
\usepackage{amsmath,amssymb,amsfonts}%
\usepackage{amsthm}%
\usepackage{mathrsfs}%
\usepackage[title]{appendix}%
\usepackage{textcomp}%
\usepackage{manyfoot}%
\usepackage{booktabs}%
\usepackage{algorithm}%
\usepackage{algorithmicx}%
\usepackage{algpseudocode}%
\usepackage{listings}%
\usepackage{ragged2e}%
\usepackage{dcolumn}%
\usepackage[dvipsnames]{xcolor}%
\usepackage{tikz}%
\usetikzlibrary{matrix}%
\usetikzlibrary{calc}%
\usetikzlibrary{fit}%
\usepackage{bm}%
\usepackage{ulem}%
\usepackage{enumerate}%

\newtheorem{definition}{Definition}
\newtheorem{theorem}{Criteria}

\newcommand{\Ocentered}{\text{$\bullet$\raisebox{0.5ex}{\rule{0.7em}{0.4pt}}$\bullet$}}

\newcommand{\SmallSquarePoints}{
    \begin{tikzpicture}
        \fill (0, -0.1) circle (0.175em);  
        \fill (0.2, -0.1) circle (0.175em);  
        \fill (0, 0.1) circle (0.175em);  
        \fill (0.2, 0.1) circle (0.175em);  
        
        \draw (0, -0.1) -- (0.2, -0.1);  
        \draw (0.2, -0.1) -- (0.2, 0.1);  
        \draw (0.2, 0.1) -- (0, 0.1);  
        \draw (0, 0.1) -- (0, -0.1);  
    \end{tikzpicture}
}

\DeclareMathOperator{\Tr}{Tr}

\newcommand{\ket}[1]{|#1\rangle}

\newcommand{\norm}[1]{\left\lVert#1\right\rVert}

\makeatletter
\providecommand{\@reinserts}{}
\makeatother

\begin{document}

\preprint{APS/123-QED}

\preprint{APS/123-QED}

\title{Entanglement routing via passive optics in CV-networks}

\author{David Fainsin}
\email{david.fainsin@lkb.upmc.fr}
\affiliation{Laboratoire Kastler Brossel, Sorbonne Universit\'e, ENS-Université PSL, CNRS, Coll\`ege de France, 4 place Jussieu, Paris F-75252, France}

\author{Antoine Debray}
\email{antoine.debray@lkb.upmc.fr}
\affiliation{Laboratoire Kastler Brossel, Sorbonne Universit\'e, ENS-Université PSL, CNRS, Coll\`ege de France, 4 place Jussieu, Paris F-75252, France}

\author{Ilya Karuseichyk}
\affiliation{Laboratoire Kastler Brossel, Sorbonne Universit\'e, ENS-Université PSL, CNRS, Coll\`ege de France, 4 place Jussieu, Paris F-75252, France}

\author{Mattia Walschaers}
\affiliation{Laboratoire Kastler Brossel, Sorbonne Universit\'e, ENS-Université PSL, CNRS, Coll\`ege de France, 4 place Jussieu, Paris F-75252, France}

\author{Valentina Parigi}
\affiliation{Laboratoire Kastler Brossel, Sorbonne Universit\'e, ENS-Université PSL, CNRS, Coll\`ege de France, 4 place Jussieu, Paris F-75252, France}

\date{\today}

\begin{abstract}
Large-scale operations in quantum networks require efficient sorting of desired paths between nodes. In this article, we consider entanglement routing, which involves establishing an entanglement link between specific nodes in a large network of bosonic nodes. The networks are continuous-variable graph states built from finite squeezing and passive linear optics, shaped by complex network structures that mimic real-world networks. We construct a bipartite routing protocol with the specific goal of establishing a teleportation channel between two clients via  passive optics operations locally operated by two different providers sharing the network. We provide criteria for extracting the aforementioned channel and, through the use of a derandomised evolutionary algorithm, extend the existing framework to study complex graph topologies.
\begin{description}
\item[Keywords]
Quantum Optics, Routing, Quantum Information, Quantum Communications, \\Continuous variables
\end{description}
\end{abstract}

\maketitle

\section*{Introduction}\label{sec:intro}

Routing is the process of defining specific paths within a network to facilitate particular operations. In quantum networks, routing entails embedding paths into the network's quantum structure. This work focuses on configuring targeted entanglement links in a continuous-variable (CV) quantum network by reshaping its entanglement structure through linear optical transformations.

In prior studies, routing has been explored extensively in qubit- or spin-based networks, either in a hardware-agnostic context \cite{Bapat23, MiguelRamiro23} or with hardware-specific approaches \cite{Paganelli13, Bottarelli23, Dutta23}. Typical qubit-based routing tasks rely on local pairwise operations, such as switching gates, e.g. swap gates \cite{Bapat21}, teleportation protocols \cite{Devulapalli24}, or complementation techniques \cite{Hahn19}, often aiming to distribute Bell pairs \cite{Beaudrap20, Meignant19}. However, studies on routing within CV quantum networks remain sparse \cite{Sansavini_2019, centrone2023cost, Zhang09}. 

In this work, we examine CV quantum networks, formally described as CV graph states \cite{Gu09}, shared between two providers. Each provider has local access to a subset of network nodes, on which they can perform linear optical transformations. Each node represents a mode of the electromagnetic field distributed to a client aiming to establish an entanglement link with a client on the opposite provider's side while being decoupled from other clients in the network. The goal is to form an EPR link between selected nodes, enabling a teleportation task. 

Unlike qubit-based networks, where graph structures are typically constructed through EPR pair distribution, CV networks benefit from the ability to generate large Gaussian graph states efficiently  \cite{Asavanant19,Larsen19}. This enables the reconfiguration of existing entanglement links within local servers to meet specific routing demands. In this context, we focus on large CV entangled networks inspired by statistical models that mimic real-world network structures. Indeed, for practical applications, the distribution of entanglement can align with the physical architecture of contemporary internet connections. Thus, this work also evaluates the efficiency of various real-world complex network structures for specialized quantum tasks.
When the number of interconnected agents in a physical system grows, it in fact tends to adopt a  structure which is neither deterministic nor random, and is described within the framework of complex networks \cite{barabasi2016network,newman2018networks}. Such networks were extensively studied over the past decades to describe real world complex networks such as the internet \cite{AS_graph}, the world-wide web \cite{BA99} or the brain \cite{brain_net} and recently to quantum technologies \cite{Nokkala2024}. 

The paper is organized as follow, we first recall some elementary notions for building graph states with continuous variables optical systems as well as a mathematical framework for entanglement routing. In a second part, we search for a solution, either ideal or approximate, to the routing problem using an efficient derandomized evolutionary algorithm.  Finally, we propose an analytical approach to the problem, providing two criteria on the extraction of a ideal teleportation channel. We extend this framework to study the possibility of routing in large complex networks topology.

\section{Results}\label{sec:graph}

\subsection{Problem statement and notation}\label{subsec:th_frm_graph}

\subsubsection{Continuous variables graph states}

Given an undirected graph $G = \{V, E\}$ of size $|V|=n$, the corresponding graph state $\ket{G}$ is a quantum state constructed by initializing each vertex in vacuum infinitely squeezed states along the momentum direction $\ket{0}_p^{\otimes n}$ and applying controlled-phase ($C_Z$) gates according to the edges of the graph. This state encapsulates the non-classical correlations inherent in the system. A key tool for studying these states is the \textit{adjacency matrix}  $A_G \in \mathcal{M}_n(\mathbb{R}) $. For a vertex set  $V = \{v_1, v_2, \dots, v_n\}$, the matrix elements $ a_{i,j}$ are defined as:  
\[
a_{i,j} =
\begin{cases} 
1 & \text{if } (v_i, v_j) \in E, \\
0 & \text{otherwise}.
\end{cases}
\]  
Since $G$ is assumed to be undirected and unweighted, $A_G$ is symmetric, reflecting equal interaction strengths between the interacting modes.

\begin{figure}
\begin{center}
\includegraphics[width=\linewidth]{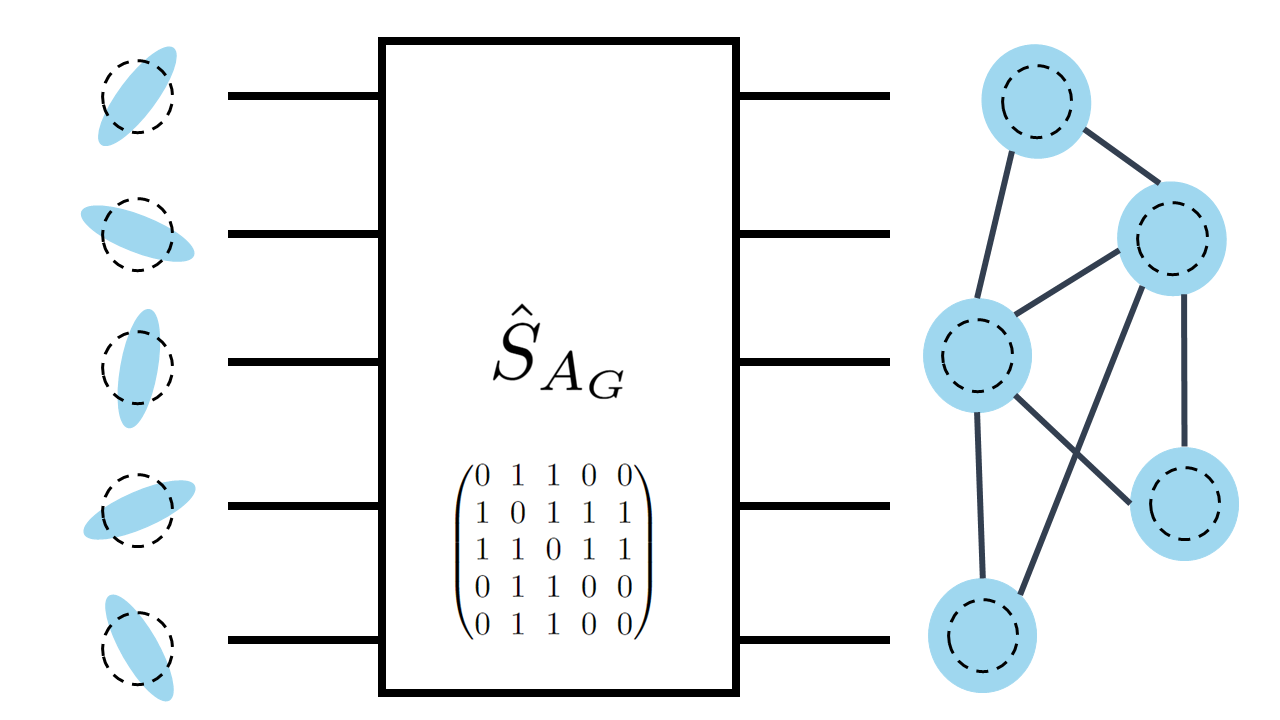}
\end{center}
	\caption{\small \textbf{Optical cluster state generation from squeezed states and linear optics}. The adjacency matrix $A_G$ is explicited in the central interferometer. The shot noise is depictured with the black dotted line.}
	\label{fig:cluster_unit}
\end{figure}

We rely on vacuum squeezed states in the momentum quadrature to obtain an approximation of a cluster state $\ket{\tilde{G}} =\hat{C}_Z(A_G)\left(\mathcal{S}(r)\ket{0}\right)^{\otimes n}$
where $\mathcal{S}(r)$ is the squeezing operator and $r$ the squeezing factor.
The implementation of the $\hat{C}_Z(A_G)$ is experimentally costly, requiring in-line squeezing. A standard approach in optics is to apply the Bloch-Messiah reduction \cite{BLOCH1962,Horoshko_2019} to the full symplectic operator $\hat{U} = \hat{C}_Z(A_G)\mathcal{S}(r)^{\otimes n}$ acting on the vacuum:

\begin{equation}
\ket{\tilde{G}} = \hat{S}_{A_G} \left[\bigotimes_{k=1}^{n} \mathcal{S}(r_k)\right] \ket{0}^{\otimes n}.
\label{eq:cluster_3}
\end{equation}
Eq. \eqref{eq:cluster_3} shows that any cluster state can be built experimentally out of a set of $n$ squeezers acting on vacuum and a unitary transformation function of the adjacency matrix of the graph $\hat{S}_{A_G}$, as shown in Fig.~\ref{fig:cluster_unit}. 
A compact analytical formula for the
orthogonal matrix $\hat{S}_{A_G}$ is defined following the associated passive unitary transformation   $U_{A_G}=X_{A_G}+iY_{A_G}$ and built as follow \cite{Ferrini_2015}
\begin{equation}
\hat{S}_{A_G} = \begin{pmatrix}
X_{A_G} & -Y_{A_G} \\ Y_{A_G} & X_{A_G}
\end{pmatrix},
\end{equation}
with 
\begin{equation}
\begin{aligned}
X_{A_G} &= \left(\bm{1} + A_G^2\right)^{-1/2}O, \\
Y_{A_G} &= A_G\left(\bm{1} + A_G^2\right)^{-1/2}O,
\end{aligned}
\label{eq:XY_cluster}
\end{equation}
where $O$ is an arbitrary orthogonal matrix, but it can be used to optimized the graph generation when the squeezing spectrum $\mathcal{S}(r_k)$ is fixed by experimental constraints \cite{Ferrini_2015}. In this work, $O$ is fixed to the identity. The impact of this choice is discussed in the last section.

Through this paper, the graph state is built with a set $n$ squeezed states with equal squeezing parameter $r_k=\sigma_0\exp(2r)$, for all $k$, with $\sigma_0$ the standard deviation of the quantum fluctuations of the vacuum. We set $\sigma_0 = 1$ (i.e. $\hbar=2$).

\subsubsection{Entanglement routing}

\begin{figure*}
\begin{center}
\includegraphics[scale=0.23]{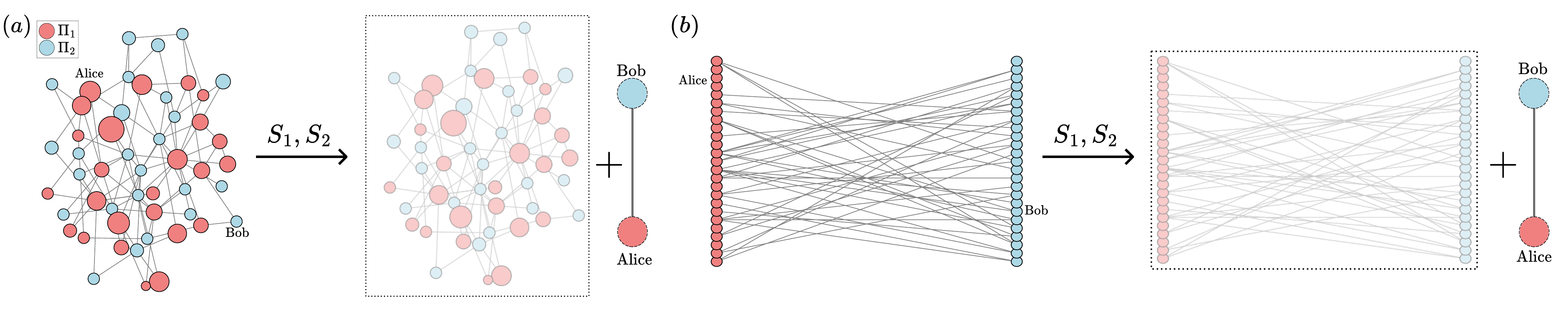}
\end{center}
	\caption{\small \textbf{Routing in a Barabasi-Albert network in a bipartite scenario}. Alice and Bob's providers are shown to share a graph state, as illustrated in (a) on the left, where each node represents a mode of the electromagnetic field.  A red node is attributed to Alice, while a blue node is attributed to Bob. In the initial figure, the dimensions of each node are directly proportional to its degree, i.e. the number of links that point toward it. Simultaneously, local passive unitary transformations are performed by Alice and Bob's providers to extract an EPR pair, as illustrated on the right side of the figure. Here the routing is done between one node of $\Pi_1$ and one of $\Pi_2$ but it could be internal targeting a pair between two nodes of a single provider. The opacity on the right hand side of both figures signifies that the resulting $n-2$ state does not necessarily constitute a graph state, no stipulations are made regarding the precise form of this state. (b) Bipartite layout of the procedure.}
	\label{fig:CN_4}
\end{figure*}

Assume two providers, $ \Pi_1 $ and $ \Pi_2 $, each control half of a quantum network of arbitrary size $ n $. These providers serve numerous clients who wish to communicate securely. The scheme is shown in Fig. \ref{fig:CN_4}. Two scenarios may arise: either both clients belong to the same provider and seek to establish a secure communication channel, or the clients belong to different providers and require inter-provider communication. The two issues can be addressed by establishing a teleportation channel between the clients.

There exists a large variety of optically multimode squeezing sources and CV entangled resources    \cite{
Larsen19, Asavanant19, Xanadu22, Kashiwazaki23, Inoue23,Kouadou_2023,barakat2024} that can be shaped and distributed to the providers. The routing approach in this work  relies on passive symplectic transformations on the mode quadratures \cite{Sansavini_2019}. While a less restrictive approach allowing for all symplectic transformations (including inline squeezing) was studied in \cite{Zhang09}, we focus here on less costly operations that can be implemented by the providers,  i.e.   passive unitaries that can be efficiently implemented via multiport interferometers \cite{William16}.


The Von Neumann Entropy plays an important role in quantum communication. It is commonly used to compute key rates in Quantum Key Distribution with continuous variables as it defines the average information that can be obtained from a quantum state measurement \cite{Neumann32,Devetak_2005}. Moreover, such quantity is invariant under unitary transformation $S(\hat{\rho}_G) = S(U\hat{\rho}_G U^\dagger)$, i.e., invariant in the context of entanglement routing with passive transformations. When $\hat{\rho}_G$ is a n-mode gaussian state, the Von Neumann Entropy can be computed from its symplectic eigenvalues $(\lambda_i)_{i=1...n}$\cite{demarie2012,Weedbrook_2012}
\begin{equation}
\label{eq: von neu}
S(\hat{\rho}) = \sum_{i=1}^{n} G \left(\frac{\lambda_i -1}{2}\right),
\end{equation}
where 
\begin{equation}
G := x \longrightarrow (x+1)\log_2(x+1) - x\log_2(x).
\end{equation}
In our context, both providers share a Graph state represented by the covariance matrix $\Gamma_G$, where $\Gamma_{Gij}=(1/2)\langle \hat{\xi}_i\hat{\xi}_j + \hat{\xi}_j\hat{\xi}_i \rangle $, being $\hat{\xi}_{i,j}$ any $q_{i,j}$ or $p_{i,j}$ quadrature of the fields belonging to the vector $\hat{\mathbf{\xi}}^T= (\hat{Q},\hat{P})^T$. $\Gamma_{G}$ can be decomposed in block matrices
\begin{equation}
    \Gamma_G = \begin{pmatrix}
        \Gamma_A & \Gamma_{AB} \\
        \Gamma_{BA} & \Gamma_B
    \end{pmatrix}.
\end{equation}
The symplectic eigenvalues will be obtained using the Williamson decomposition on $\Gamma_A$ (and $\Gamma_B$).
The Williamson decomposition of $\Gamma_A$ is then
\begin{equation}\label{eq:williamson}
    \Gamma_A = S_A \Gamma_W S_A^T
\end{equation}
with $S_A$ a symplectic transformation and $\Gamma_W = \text{diag}(\lambda_1,..,\lambda_n)$.

\subsubsection{Target state}\label{subsubsec:target}

As aforementioned, two clients, Alice and Bob, aim to establish an entangled bipartite state with the highest possible Von Neumann entropy and the two providers must configure the network to achieve it. Such state can be a CV EPR state characterized by its variances $\Delta^2 (\hat{q}_a-\hat{q}_b)=2e^{-2r}$ and $\Delta^2 (\hat{p}_a+\hat{p}_b)=2e^{-2r}$, being $e^{-2r}$ a squeezing factor.
To align with the graph-state framework, we define our practical target state ($\Ocentered$) as a two-mode linear cluster state, which can be implemented using linear optics.
The adjacency matrix of this two-mode linear cluster state is trivial, reading
\begin{equation}
A_{\Ocentered} = \begin{pmatrix}
0 & 1 \\
1 & 0
\end{pmatrix}
\end{equation}
and the transformation needed to implement on the momentum-squeezed modes to obtain such a cluster is
\begin{equation}
S_{\Ocentered} = \begin{pmatrix}
X_{\Ocentered} & -Y_{\Ocentered}\\
Y_{\Ocentered} & X_{\Ocentered}
\end{pmatrix} = \frac{1}{\sqrt{2}}\begin{pmatrix} 
1 & 0 & 0 & -1\\
0 & 1 & -1 & 0\\
0 & 1 & 1 & 0 \\
1 & 0 & 0 & 1
\end{pmatrix}
\end{equation}
where we recall that $X_{\Ocentered}$ and $Y_{\Ocentered}$ are linked to $A_{\Ocentered}$ by the eq. (\ref{eq:XY_cluster}). Moreover, in our case the parameter $O$ is fixed to identity since we assume identical squeezing in all modes.

The covariance matrix of the resulting finite-squeezed cluster state is given by $\Gamma_{\Ocentered} = S_{\Ocentered}\Gamma_{\text{sqz}}S_{\Ocentered}^\dagger$ where $\Gamma_{\text{sqz}}= diag(e^{2r},e^{2r},e^{-2r},e^{-2r})$ is the diagonal covariance matrix for the squeezed states.
Overall the target state's covariance matrix is
\begin{equation}
\Gamma_{\Ocentered} =\begin{pmatrix}
\lambda & 0 & 0 & \mu \\
0 & \lambda & \mu & 0 \\
0 & \mu & \lambda & 0 \\
\mu & 0 & 0 & \lambda
\end{pmatrix}
\end{equation}
with $\lambda = \cosh (2r)$, $\mu = \sinh (2r)$. $\Gamma_{\Ocentered}$ exhibits similar correlation strength as $\Gamma_{EPR}$, with correlations between conjugate quadratures of the two modes, rather than direct correlations within the same type of quadrature (i.e. position-position or momentum-momentum). From an information theory point of view these states can be considered as equivalent and $\Gamma_{\Ocentered}$ can be used equivalently to perform a teleportation protocol simply by adjusting the feedforward classical post-processing done between Alice and Bob \cite{Pirandola_2006}. Such a state is maximally entangled and can be used as a teleportation channel between the two clients Alice and Bob.


Let us define the initial graph state $\ket{G_i}$. Our objective is to route this initial graph state to a target state $\ket{G_t}$. It can be demonstrated that if a complete access to all the nodes of the initial state is provided, a complete rewriting of the state is feasible and it is possible to reach any state $\ket{G_t}$. Applying a global transformation on a cluster to switch from $\ket{G_i}$ to $\ket{G_t}$ is nearly trivial knowing how to build them via symplectic transformation, $S_i$ and $S_t$, on vacuum squeezed states $(\hat{Q}_s,\hat{P}_s)$.
\begin{equation}
\begin{matrix}
\begin{pmatrix}
    \hat{Q}_c^i \\ \hat{P}_c^i
    \end{pmatrix} = S_i\begin{pmatrix}
    \hat{Q}_s \\ \hat{P}_s
\end{pmatrix}
&
\begin{pmatrix}
    \hat{Q}_c^t \\ \hat{P}_c^t
    \end{pmatrix} = S_t\begin{pmatrix}
    \hat{Q}_s \\ \hat{P}_s
    \end{pmatrix}
\end{matrix}
\label{eq:R1}
\end{equation}
The symplectic group structure allow then us to define $S_{i\rightarrow t} = S_t S_i^{-1}$ and thus
\begin{equation}
\begin{pmatrix}
    \hat{Q}_c^t \\ \hat{P}_c^t
    \end{pmatrix} = S_{i\rightarrow t} \begin{pmatrix}
    \hat{Q}_c^i \\ \hat{P}_c^i
    \end{pmatrix}.
\end{equation}
Thus, there is a straightforward solution to the routing problem when a global transformation can be applied. As such, in the case where Alice and Bob are two clients of a unique provider, the routing problem is straightforward, as a complete access to the graph is possible.

In a more realistic communication scenario the modes of the cluster are distributed to two spatially separated providers, $\Pi_1$ and $\Pi_2$, such that each provider is allowed to perform local linear optical transformations only on its set of nodes. Say $n_1$ and $n_2$ are the number of nodes of the providers of Alice and Bob respectively such that $n=n_1+n_2$. The application of local linear optical transformations for both providers, designated as $S_{1}$ and $S_{2}$, to the respective $n_1$ and $n_2$ modes results in the following global symplectic transformation of the initial state

\begin{equation}
S=\begin{pmatrix}
    Re(U_1) & \mathbf{0}_{n_1 n_2} &-Im(U_1)& \mathbf{0}_{n_1 n_2}\\
    \mathbf{0}_{n_2 n_1}&Re(U_2)&\mathbf{0}_{n_2 n_1}&-Im(U_2)\\
    Im(U_1)&\mathbf{0}_{n_1 n_2}&Re(U_1)&\mathbf{0}_{n_1 n_2}\\
    \mathbf{0}_{n_2 n_1}&Im(U_2)&\mathbf{0}_{n_2 n_1}&Re(U_2)
    \end{pmatrix},
\label{eq:R4}   
\end{equation}
where $U_1$ and $U_1$ are the unitary matrices associated to $S_{1}$ and $S_{2}$ and parametrized respectively by $n_1^2$ and $n_2^2$ parameters. The null matrices ($\mathbf{0}_{n_1 n_2}$ and $\mathbf{0}_{n_2 n_1}$) probe the constraints imposed by local transformations on the routing problem, demonstrating that certain changes are inaccessible and limiting the available degrees of freedom.

The primary objective of this work is to achieve ideal routing, as defined below, using passive linear optics. We simulate this process using a derandomised evolution strategy, as talked about in the next section, to identify which graph type allows such a transformation.

\begin{definition}[\textit{Ideal routing}]
    The concept of \textbf{Ideal routing} is defined as the ability to reach a state $\ket{G_t}$ in which a maximally entangled EPR pair is shared between the two desired parties, such a state as a maximal Von Neumann Entropy.
\end{definition}
Should this prove unsuccessful, we will explore the possibility of achieving imperfect routing.
\begin{definition}[\textit{Imperfect routing}]
   The concept of \textbf{imperfect routing} is defined as the ability to reach a state of the system, denoted by $\ket{G_t}$, in which a pair of modes is shared between two parties, with the condition that this shared pair is entirely disentangled for the remainder of the graph.
\end{definition}

\subsection{Entanglement Routing via derandomized evolution strategy}\label{sec:algo}


A first approach to find the appropriate unitaries $(U_1 U_2)$ for the providers to establish a teleportation channel between modes of Alice and Bob (noted $m_A$ and $m_B$) is done through a feedback loop, using a robust evolutionary algorithm capable of scanning a large parameter space. All these parameters together permit the generation of unitary matrices in a deterministic way, see Appendix~\ref{sec:Gellmann} on Gell-Mann parametrization of unitaries for more. 

To design a robust evolutionary algorithm, we follow a derandomized evolution strategy (DES) as it has already demonstrated efficient behavior when dealing with Covariance Matrix Adaptation \cite{Hansen01,Roslund09}. DES are similar to genetic methods; that is, the way parameters are updated is not purely stochastic and depends on the quality of the existing offspring to realize the mutations, crossover, and selection. More precisely, we used the Covariance Matrix Adaptation Evolutionary Algorithm (CMA-ES), the algorithm's operation relies on the adaptive update of a covariance matrix, capturing statistical relationships among variables in a group of candidate solutions. This covariance matrix models the multivariate distribution of the population and is iteratively adjusted to converge toward the optimal solution. A detailed description of the CMA-ES algorithm is provided in Appendix~\ref{sec:AppA} and is illustrated in Figure~\ref{fig:CMA_algo}. 

\begin{figure}
\begin{center}
\includegraphics[width=\linewidth]{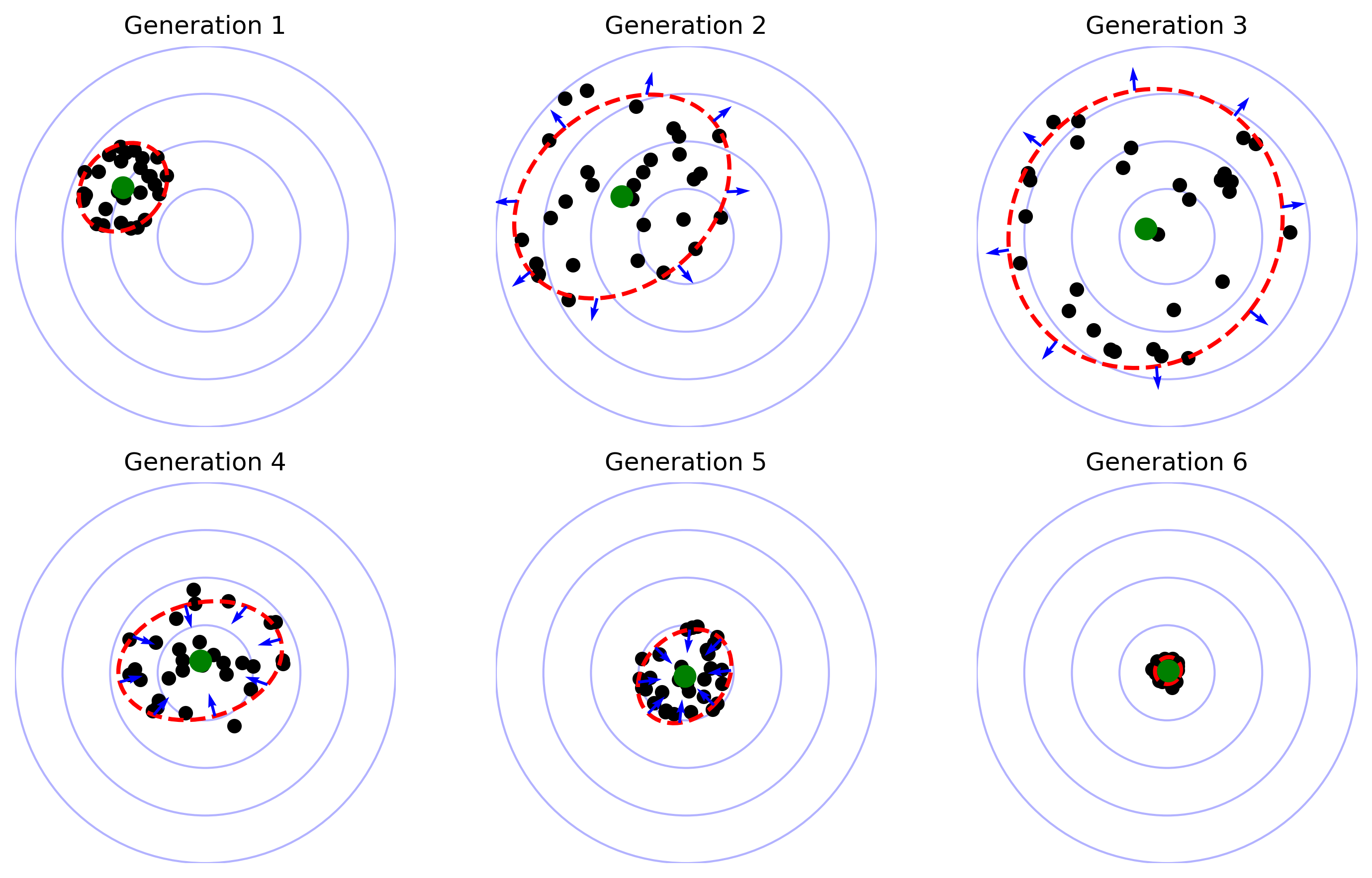}
\end{center}
	\caption{\small \textbf{Illustration of the CMA-ES algorithmic process}. In green the center of the ellipse. The blue arrows show the expansion and contraction of the area explored, the algorithm begins by exploring a large diversity of offspring, to limit the influence of local minima and then progressively reduces the diversity to look for the precise solution.}
	\label{fig:CMA_algo}
\end{figure}
The most successful offsprings guide the search towards promising areas of the search space. To evaluate our different offsprings we make use of the Froebenius norm ($\norm{\cdot}_2$) between their associated covariance matrix and the ideal covariance matrix targeted. 
We define the \textbf{ideal covariance matrix} $\Gamma_{\text{ideal}}(m_A, m_B)$ as a generalization of $\Gamma_{\Ocentered}$ between nodes  $m_A$ and $m_B$. This matrix, with dimensions $4 \times 2n$, encapsulates the desired correlations between the nodes, ensuring that the correlations resemble those of an EPR pair and no correlations with the other modes.
Thus, we work with the \textbf{reduced covariance matrix} $\Gamma_{\text{reduced}}(m_A, m_B)$, which captures only the relevant correlations between the selected nodes while ignoring the others.
\begin{widetext}
\[ \Gamma_G = \begin{tikzpicture}[baseline=(math-axis),every right delimiter/.style={xshift=-3pt},every left delimiter/.style={xshift=3pt}]%
\matrix [matrix of math nodes,left delimiter=(,right delimiter=)] (matrix)
{
|(m11)| a_{0,0} & |(m12)| \cdots & |(m13)| a_{0,m_A} & |(m14)| a_{0,m_B} & |(m15)| a_{0,m_A+n} & |(m16)| a_{0,m_B+n} & |(m17)| \cdots & |(m18)| a_{0,2n-1} \\
|(m21)| \vdots & |(m22)| \cdots & |(m23)| & |(m24)| & |(m25)| & |(m26)| & |(m27)| \cdots & |(m28)| \vdots \\
|(n11)| a_{m_A,0} & |(n12)| \cdots & |(n13)| a_{m_A,m_A} & |(n14)| a_{m_A,m_A} & |(n15)| a_{m_A,m_A+n} & |(n16)| a_{m_A,m_B+n} & |(n17)| \cdots & |(n18)|a_{m_A,2n-1}\\ 
|(n21)| a_{m_B,0} & |(n22)| \cdots & |(n23)| a_{m_B,m_A} & |(n24)|  & |(n25)|  & |(n26)| a_{m_B,m_B+n} & |(n27)| \cdots & |(n28)|a_{m_B,2n-1}\\ 
|(n31)| a_{m_A+n,0} & |(n32)| \cdots & |(n33)| a_{m_A+n,m_A} & |(n34)|  & |(n35)|  & |(n36)| a_{m_A+n,m_B+n} & |(n37)| \cdots & |(n38)|a_{m_A+n,2n-1}\\ 
|(n41)| a_{m_B+n,0} & |(n42)| \cdots & |(n43)| a_{m_B+n,m_B} & |(n44)| a_{m_B+n,m_B} & |(n45)| a_{m_B+n,m_A+n} & |(n46)| a_{m_B+n,m_B+n} & |(n47)| \cdots & |(n48)|a_{m_B+n,2n-1}\\ 
|(m31)| \vdots & |(m32)| \cdots & |(m33)| & |(m34)| & |(m35)| & |(m36)| & |(m37)| \cdots & |(m38)| \vdots \\
|(m41)| a_{2n-1,0} & |(m42)| \cdots & |(m43)| a_{2n-1,m_A} & |(m44)| a_{2n-1,m_B} & |(m45)| a_{2n-1,m_A+n} & |(m46)| a_{2n-1,m_B+n} & |(m47)| \cdots & |(m48)| a_{2n-1,2n-1}\\
};
\node[draw,dashed,color=blue,inner sep=0pt,fit=(n11) (n18) (n48) (n41) (n11)] {};
\node[draw,dashed,color=red,inner sep=0pt,fit=(n13) (n16) (n46) (n43) (n13)] {};
\coordinate (math-axis) at ($(matrix.center)+(0em,-0.25em)$);
\node[draw,color=white,text=red] at ($(matrix.center)+(-1em,+0.25em)$) {$\Gamma_{routed}$};
\node[draw,color=white,text=blue] at ($(matrix.center)+(-1em,+3.5em)$) {$\Gamma_{reduced}$};
\end{tikzpicture}\]
\end{widetext}

We then compare the reduced covariance matrix generated by tracing out all lines of the covariance matrix built from the offspring except those associated to nodes $m_A$ and $m_B$. The quantity
\begin{align*}
    \norm{\Gamma_{\text{ideal}}(m_A,m_B) -\Gamma_{\text{reduced}}(m_A,m_B)}_2
\end{align*}
should be minimized in our optimization problem, and ideally converge to zero in a successful routing scenario. However with real-world complex networks, the extraction of a perfect teleportation channel is not always possible, and the algorithm is sensitive to local minima. To remedy this problem we optimize the state purity $\gamma$. The covariance matrix to consider is the $4\times 4$ matrix of the two nodes of interest $(m_A,m_B)$ obtained when tracing out all the other modes. We call this matrix $\Gamma_{\text{routed}}$. In the case of a successful routing $\Gamma_{\text{routed}}$ should be identical to $\Gamma_{\Ocentered}$. These three matrices allow the evaluation of the offspring quality by minimizing the following objective function
\begin{equation}\label{eq:R18}
\begin{aligned}
    f_{opt} = & \norm{\Gamma_{\text{ideal}}(m_A,m_B) -\Gamma_{\text{reduced}}(m_A,m_B)}_2 + \\
     & \frac{1}{2}\left[1-\gamma(\Gamma_{\text{routed}}(m_A,m_B))\right].
\end{aligned}
\end{equation}

\begin{figure*}
\begin{center}
\includegraphics[width = \textwidth]{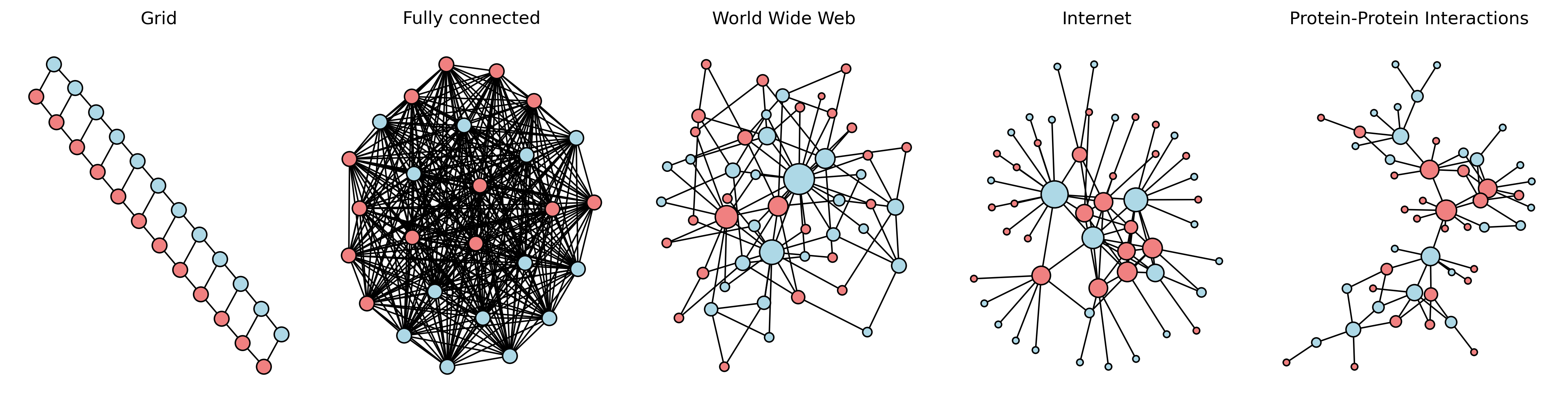}
\end{center}
	\caption{\small \textbf{Deterministic and complex topologies used in this paper}. The graphs were generated using the Networkx library. Different colors are associated to different providers.}
	\label{fig:net_ex}
\end{figure*}

We now present the most recent results obtained based on the evolutionary algorithm. We explored several different topologies, ranging from deterministic to complex networks, presented in Figure~\ref{fig:net_ex}. For computation complexity reasons, we limit our study to graphs with 50 nodes maximum, which already captures significative differences in quantum networks drawn via complex networks models \cite{Renault2023}.  The detailed code can by found in \cite{Fainsin_Entanglement_Routing_Simulation}.

\subsubsection{On deterministic graphs}

For deterministc graphs some results have been obtained in \cite{Sansavini_2019} for small regular networks. In the case of fully connected networks, one can notice a dichotomy between internal and bi-partite routing (i.e. routing between two nodes of the same provider and routing between the nodes of two distinct providers). Similarly, in the case of grid-type networks, a dichotomy occurs depending on the parity of the nodes possessed by each provider. The total number of nodes in the graph is given by $n=2k$, where $k$ is the number of nodes belonging to each provider. The observed behaviours vary in accordance with the parity of $k$. We call Even Grid (resp. Odd Grid) the case when $k$ is even (resp. odd). Those results are summarized in Table~\ref{tab:res_fra}. 

\begin{table}[h!]
\centering
    \begin{tabular}{|c|c|c|c|}
        \hline
        Topology & Even Grid & Odd Grid & Fully Conn.  \\
        \hline
        Routing (Alice/Bob) & No & Yes & No \\
        Routing (Alice/Alice) & No & No & Yes\\
		\hline        
    \end{tabular}
    \caption{\textbf{Results overview for deterministic graphs}}
    \label{tab:res_fra}
\end{table}

We have confirmed this behaviour for all possible grids of sizes up to 100 nodes, in the cases where routing is possible the objective function tend to converge to zero. The parameters obtained permits to build the unitaries for both parties (Alice and Bob) such that $\Gamma_{routed}=\Gamma_{\Ocentered}$. On the other hand, when routing is not possible, the objective function tends to quickly stop progressing and reach some local extrema with $f_{opt}=\mathcal{O}(10^{-1})$. Let us note that the difference between behaviors decreases with the greater the size of the network, for a constant number of generations, but  the difference in convergence speed  persists. In this situation, $\Gamma_{routed}$ has a specific shape that we sum up in the following way: 

\begin{equation}
\Gamma_{\text{imperfect, AB}} = \begin{pmatrix}
\lambda -\epsilon & \mathcal{O}(10^{-9}) & \mathcal{O}(\epsilon) & \mu' \\
\mathcal{O}(10^{-9}) & \lambda -\epsilon  & \mu' & \mathcal{O}(\epsilon) \\
\mathcal{O}(\epsilon) & \mu' & \lambda +\epsilon  & \mathcal{O}(10^{-9})\\
\mu' & \mathcal{O}(\epsilon) & \mathcal{O}(10^{-9}) & \lambda +\epsilon 
\end{pmatrix}
\label{eq:not_converge_grid}
\end{equation}
which still exhibit undesired correlation at the end of the execution (see $\mathcal{O}(\epsilon)$). Yet $\gamma\left(\Gamma_{\text{imperfect, AB}}\right) = 1$, thus the state is decorrelated from the rest of the network but does not exhibit the exact correlations of a the desired target state. However, it should be noted that this imperfect state is entirely decoupled from the rest of the graph. While it does not manifest the correlations of a EPR pair, it may exhibit two-way steering and could still be used as a communication channel. 

\subsubsection{On complex topologies}

The implementation of complex clusters in our problem allows to study models that closely resemble real networks. We examine the protocol for extracting $\Gamma_{\Ocentered}$ on different complex topology following construction rules established by the Barabási-Albert (BA) model, internet as a graph (AS) model and a duplication-divergence (PP) model. However, with these complex topologies the graph state exhibits new properties such as small worlds and hubs. Thus we run our algorithm also on a diverse range of scenario where the routing algorithm is performed on three distinct cases: 
\begin{enumerate}[I]
\item Alice is a hub node and Bob a node with low degree;
\item Alice and Bob are two nodes with maximum distance in the graph;
\item Alice and Bob are two nodes with low degrees.
\end{enumerate}
In total, for each topologies and each scenario we run the algorithm 100 times on graph of size $n=50$ through 20 000 generations. The results are summed up in Table.~\ref{tab:res_algo}.

\begin{widetext}
\begin{center}
\begin{table}[h!]
\centering
    \begin{tabular}{|c||c|c|c||c|c|c||c|c|c|}
\hline
& \multicolumn{3}{|c||}{BA} & \multicolumn{3}{|c||}{AS} & \multicolumn{3}{|c|}{PP}\\
\hline
Scenario                             & I & II & III & I & II & III & I & II & III\\
\hline
Purity $\bar{\gamma} ~(\%)$          & 97.49 & 98.43 & 98.83 & 98.97 & 99.54 & 99.34 & 98.15 & 98.45 & 98.04 \\
        $\text{std}(\gamma) ~(\%)$   & 4.07 & 2.06 & 1.52 & 2.05 & 0.8 & 0.78 & 2.9 & 2.02 & 2.55 \\
        $\bar{f}_{opt}$              & 0.64 & 0.62 & 0.59 & 0.7 & 0.71 & 0.74 & 0.6 & 0.63 & 0.62 \\
        $\text{std}(f_{opt})$        & 0.27 & 0.23 & 0.21 & 0.38 & 0.48 & 0.45 & 0.33 & 0.32 & 0.34 \\
		\hline        
    \end{tabular}
    \caption{\textbf{Performance Summary on Complex Networks by the Algorithm}. $\bar{\gamma}$ is the average purity obtained over the 100 simulations and $\text{std}(\gamma)$ the standard deviation. Similar conventions are adopted for the objective function $\bar{f}_{opt}$ and $\text{std}(f_{opt})$. }
    \label{tab:res_algo}
\end{table}
\end{center}
\end{widetext}

The key metrics assessed are the average purity ($\bar{\gamma}$) and the optimal objective function ($\bar{f}_{opt}$), along with their respective standard deviations. The results demonstrate that the algorithm consistently achieves high purity, with values ranging from 97.49\% to 99.54\%, and greater stability (lower standard deviations) in more advanced scenarios, particularly for the AS topology. Additionally, the objective function values remain relatively stable, with improvements observed in scenarios II and III. Nonetheless, while increasing the number of generation permits to progressively improve the purity value closer to 1, $f_{opt}$ is almost never reaching values significantly close to 0 (below $10^{-5}$ considering the order of magnitudes at stake in our numerical simulations). The unitary transformation computed by the algorithm can only lead to routed state of shape $\Gamma_{\text{imperfect, AB}}$ explicited in Eq.~(\ref{eq:not_converge_grid}). Overall, these results highlight the algorithm’s robustness and effectiveness in managing complex graph states for entanglement routing tasks. The reasons behind the impossibility to perform a perfect entanglement routing with passive optical transformations is deeper explored in the next section.

\subsection{Analytical approach to Entanglement Routing}\label{sec:analytical}

\subsubsection{A simple case of routing impossibility: the square network}

To acquire a little bit of intuition on the general problem we begin with the first non-trivial case of routing where only 4 modes here shared between the providers of Alice and Bob with a square shape (\SmallSquarePoints). The passive transformation on each side is a general beam splitter combined with phase rotation meaning that the number of parameters permits to look for a manual optimization, detailed calculation is given in appendix. Following this transformation, we expect to obtain at least one decoupled EPR pair between two shared nodes.

The square network is represented by the adjacency matrix 
\begin{equation}
A_{\SmallSquarePoints} = \begin{pmatrix}
0&1&1&0\\
1&0&0&1\\
1&0&0&1\\
0&1&1&0
\end{pmatrix}.
\end{equation}
Following the construction rules of Section \ref{sec:graph}, we build the symplectic matrix $S_{\SmallSquarePoints}$. Thus, the initial covariance matrix of the state is
\begin{equation}
\Gamma_{\SmallSquarePoints} = S_{\SmallSquarePoints} \Gamma_{\text{sqz}} S_{\SmallSquarePoints}^T .
\end{equation}
The analytical form of $\Gamma_{\SmallSquarePoints}$ can be found in Appendix.~\ref{sec:AppC}. The transformations performed by each providers on their respective subsystems are governed by four parameters: $\phi_{1A}, \phi_{2A}, \phi_{3A}, \theta_A$ (and likewise $\phi_{1B}, \phi_{2B}, \phi_{3B}, \theta_B$)\cite{Leonhardt98}. We denote this set of parameters as $\mathfrak{X}$ for the sake of brevity. The unitary matrices associated with these transformations are the product of a rotation matrix and a phase-shift matrix, which allows us to study the situation in the most general way
\begin{equation}
U_{A/B} = \begin{pmatrix}
e^{i\phi_{2A/B}}\cos(\theta_{A/B}) & -e^{i(\phi_{1A/B}+\phi_{2A/B})}\sin(\theta_{A/B}) \\
e^{i\phi_{3A/B}}\sin(\theta_{A/B}) & e^{i(\phi_{1A/B}+\phi_{3A/B})}\cos(\theta_{A/B})
\end{pmatrix}.
\end{equation}
From $U_A$ and $U_B$ we build the overall simplectic transformation on the state
\begin{equation}
S = \begin{pmatrix}
S_A & 0 \\
0 & S_B 
\end{pmatrix}.
\end{equation}
It leads to the final cluster state covariance matrix in its most general form
\begin{equation}
\Gamma_{gen} = S \Gamma_{\SmallSquarePoints} S^T .
\end{equation}
The problem thus reduces to finding a solution to the equation $\Tr_{A_iB_j, i,j\in(1,2)}(\Gamma_{gen}) = \Gamma_{\Ocentered}$ for the set of parameters.
We analytically search for values of $\mathfrak{X}$  for which  $\Gamma_{gen}$ is zero in positions if $\Gamma_{\Ocentered}$ is zero. In other words, we are looking for values of $\mathfrak{X}$ such that
\begin{equation}
(\Gamma_{gen})_{i,j}(\mathfrak{X}) = 0 ~~ \text{iff} ~~ (\Gamma_{\Ocentered})_{i,j} = 0 
\label{eq:ana_pb}
\end{equation}
The details of the computation is given in Appendix. \ref{sec:AppC}. We find that there are no values of $(\phi_{1A},\phi_{2A})$ for which all wanted covariances vanish simultaneously. Overall, this study establishes that in the case of a square cluster, there is no analytical solution to the problem of ideal entanglement routing,  it is never possible with passive linear optics. This also confirms that if our evolutionary strategy does not converge perfectly, it is indeed due to an underlying physical limitation. However, this analytical result cannot be generalized yet to all cluster topologies with $n>4$.

\subsubsection{No-go criteria: the symplectic distribution of deterministic and complex topologies}

Following their local symplectic transformation and assuming they have succeeded in reaching a state $\ket{G_t}$, in which they share \Ocentered, each provider observes locally a thermal state with photon number $\lambda=\cosh(2r)$. This state is completely disentangled from the rest of all other local nodes as it is maximally entangled.

If the transformation applied by each provider are local symplectic transformations $S_{A}$ and $S_{B}$ corresponding to the Williamson decomposition of their local state, then the resulting state is locally equivalent to a set of thermal states. Consequently, an evaluation of the symplectic eigenvalues of the local states allows us to derive two no-go criteria for the routing problem. 

\begin{theorem}[\textit{Entanglement Routing}] \label{obs1}
Let $\sigma(\Gamma_W)$ be the spectrum of $\Gamma_W$, as defined in Eq.~\eqref{eq:williamson}. The spectrum $\sigma(\Gamma_W)$ is conserved under all symplectic transformation, which include passive transformations. We can thus state, if $\lambda \notin \sigma(\Gamma_W)$, then \textbf{ideal routing} is impossible.
\end{theorem}
\begin{theorem}[\textit{Internal Entanglement Routing}] \label{obs2}
In the case of internal routing a necessary condition for \textbf{ideal routing} is to have the value 1 with multiplicity 2 in the symplectic spectrum $\sigma(\Gamma_W)$.
\end{theorem}
As such, for both criteria, the presence of the right value in the symplectic spectrum is a necessary condition for ideal routing. Moreover, the symplectic values 1 indicate that some modes of the providers are pure. Thus, after symplectic transformation, the provider is free to perform global transformations on his set of pure modes. 

Criteria \ref{obs1} and \ref{obs2} are now applied in order to explore the symplectic values of several bipartite graph states with the different topologies presented in Figure \ref{fig:net_ex}. In light of the aforementioned criteria, we can explain the behaviour observed with our evolutionnary algorithm in Section~\ref{sec:algo}. Specifically, in the case of a fully connected network, we find a symplectic value of $1$ with multiplicity $n_1-1$. Furthermore, regardless of the size of the network, the last symplectic value is never equal to $\lambda$. Thus,  in this topology, routing is possible between two nodes belonging to a single provider, but not between two nodes shared by two distinct providers. Moreover, in the case of Grid networks, the symplectic spectrum never present the value 1. However, the value $\lambda$ is present only if $n_1$ is odd i.e. Alice and Bob's providers hold an odd number of modes.

\begin{figure}
\begin{center}
\includegraphics[width=\linewidth]{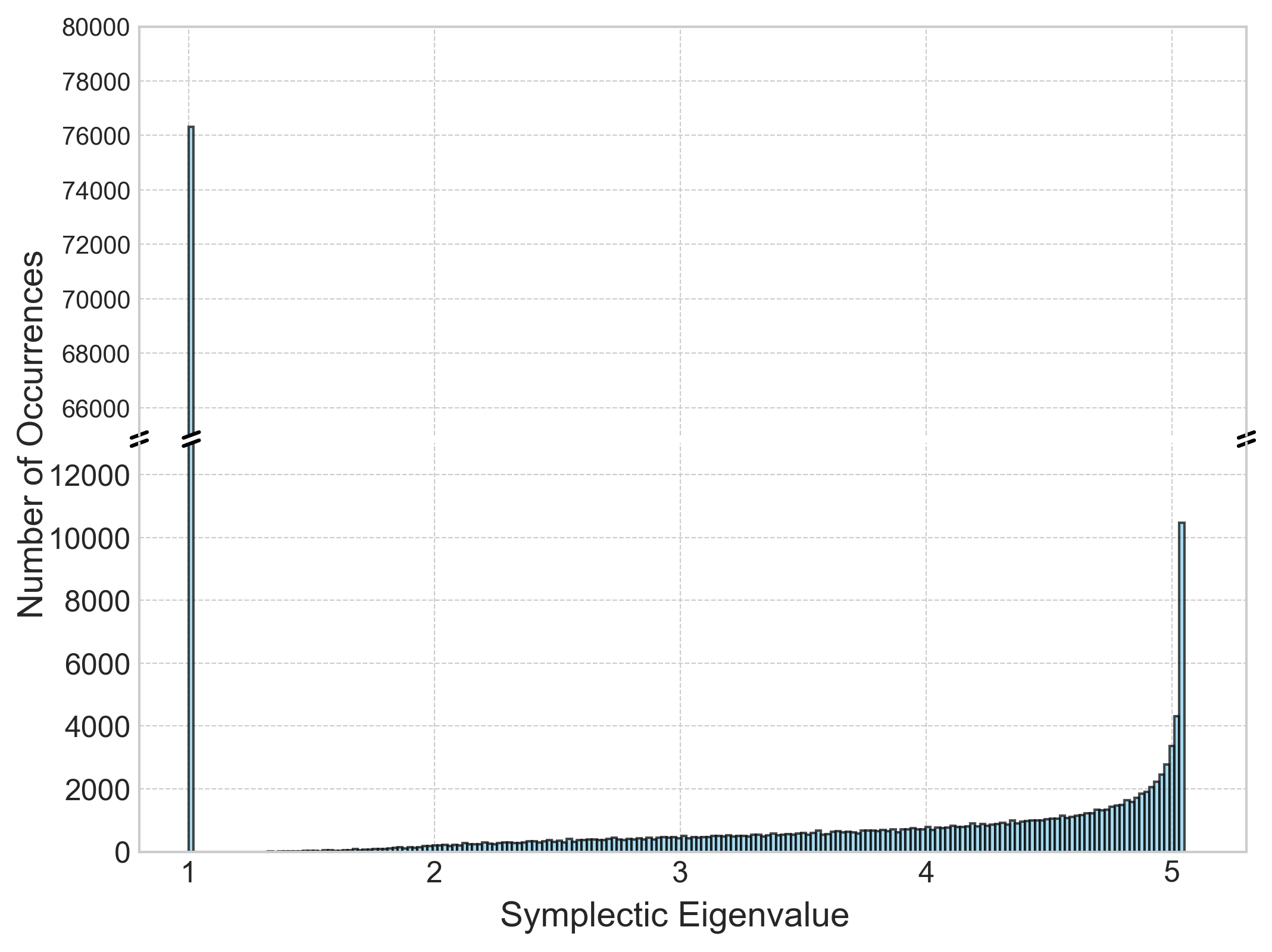}
\end{center}
	\caption{\small \textbf{Histogram of symplectic eigenvalues from 100 Barabasi-Albert graphs.} Each graph was composed of 1000 nodes equally splitted between two users leading to 200 000 symplectic values. The initial squeezing parameter is arbitrary chosen to be $s=10$ such that $\lambda=5.05$.}
	\label{fig:hist_ba}
\end{figure}

The complex topologies, on the other hand, are not built deterministically. They are studied through the distribution of their symplectic eigenvalues. We generate 100 graphs of a chosen topology with $n=1000$ and study the histogram of all the obtained symplectic values, see Figure~\ref{fig:hist_ba}. 

We arbitrarily selected $s=10$ such that $\lambda=5.05$. We sum up the result in Table \ref{tab:res_hist}. The results demonstrate that in all complex topologies, the symplectic value 1 is present with average multiplicity exceeding 30 percent of the total number of nodes, reaching even more than 75\% in internet-as-graph topologies. Therefore, all complex networks studied appears suitable for internal routing. 

\begin{table}
\centering
    \begin{tabular}{|c|c|c|c|}
        \hline
        Topology & BA & AS & PP \\
        \hline
        Value 1 (\%) & 38.158 & 77.301 & 54.759 \\
        Value $\lambda$ (\%) & 0.034 & 0.027 & 0.12 \\
        Values $\geq 99\lambda/100$ (\%) & 8.246 & 4.610 & 6.931 \\
		\hline        
    \end{tabular}
    \caption{\textbf{Occurence of the different symplectic values of interest}.}
    \label{tab:res_hist}
\end{table}

When the target state is a bipartite EPR pair, the probability to see the exact value $\lambda$ in the spectrum is much less important. For some reasons however, it is more likely to happen in a duplication divergence graphs (PP) than in the two other topologies by a significant order of magnitude, as shown in Table \ref{tab:res_hist}. Yet, as shown in Figure ~\ref{fig:hist_ba}, there is still a noticeable number of symplectic values very close to $\lambda$, and quantified in Table \ref{tab:res_hist}. The thermal state held by each providers is then of the shape of $\Gamma_{\text{imperfect, AB}}$. To wrap up, perfect routing is theoretically feasible in non-deterministic graph states as the value $\lambda$ can appear in the spectrum with low probability. But, generally, only imperfect routing is attainable. 
\begin{figure*}
\begin{center}
\includegraphics[scale=0.35]{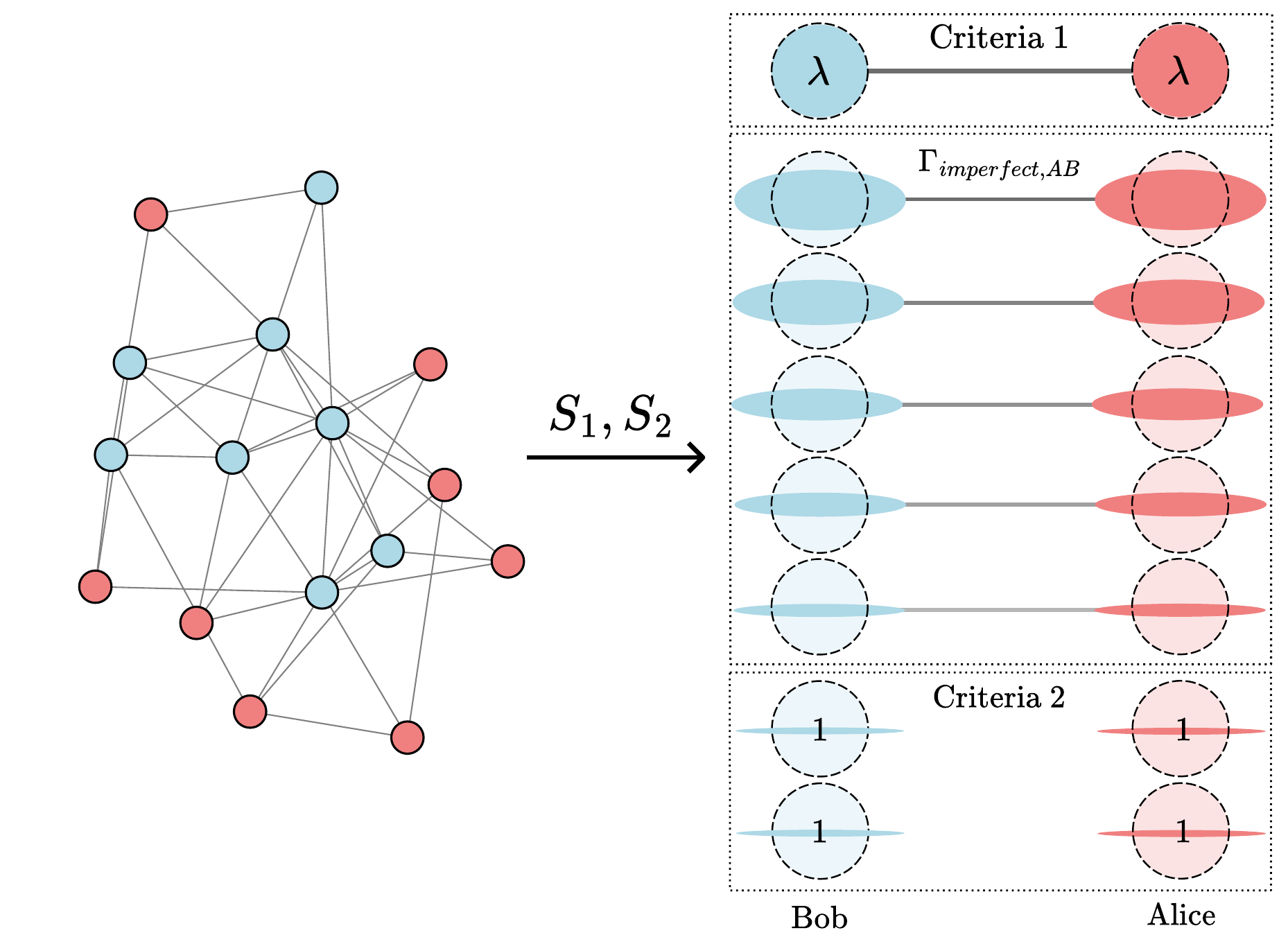}
\end{center}
	\caption{\small \textbf{Simplified representation of a bipartite graph state}. Left: a Barabasi-Albert graph with $n=16$ nodes shared between two providers. Right: the graph state after passive unitary transformation represented with a bipartite layout. The link width are proportionnal to the Von Neumann Entropy of each pair. Criteria 1 and 2 are represented. If criteria 1 is fulfilled then, as wished, an EPR pair is extracted from the graph. If criteria 2 is fulfilled for internal routing, then up two a beam-splitter transformation, a EPR pair can be created between two nodes of the same provider. If neither are fulfilled, then a number of imperfect pairs are created.}
	\label{fig:anafin}
\end{figure*}
\section{Discussion}
In light of the results, most quantum networks with complex shapes derived from real-world networks are compatible with a bipartite entanglement routing task.
Following works can determine the potential impact of the orthogonal matrix $O$ in equation \eqref{eq:XY_cluster} used in building the network structure. While not changing the shape of the network, by departing from the Identity matrix, it can be used to  optimize the squeezing distribution in the network for specific tasks \cite{Sansavini_2019,Ferrini_2015} and so have an impact in the entanglement routing problem.

Moreover, analyzing the symplectic transformation $S_A$ obtained from the Williamson decomposition will provide insight.
Applying the Bloch Messiah decomposition to $S_A$ leads to two orthogonal matrices $O_{A1}$, $O_{A2}$ and a squeezing matrix $D$ hence
\begin{equation}
\label{eq:BMWilliam}
  \Gamma_A = O_{A1} D O_{A2} \Gamma_W O_{A2}^T D O_{A1}^T .
\end{equation}
Recalling that $\Gamma_W$ represents a set of thermal states, with variance greater or equal to that of the vacuum, it shows no logarithmic negativity \cite{Walschaers21}. Consequently, $O_{A2} \Gamma_W O_{A2}^T$ and the set of squeezers $D$ creates no quantum correlations (only classical correlations) between Alice's provider nodes. Only $O_{A1}$ is responsible for the generation of entanglement within the system.
It finally implies that, in a scenario where the graph state is built with equally squeezed states, the passive transformation $U = diag(O_{A1},O_{B1})$ is a solution to the routing problem, i.e. $U^T \Gamma_G U$ is a perfect routed state if $\lambda \in \sigma(\Gamma_W)$. Moreover, in a scenario where Criteria 1 is fullfiled, it is worth noting that the associated squeezing in the matrix D is 1 (no squeezing necessary) and reversely when Criteria 2 is fullfiled the squeezing necessary to produce the original state is exactly s.

Finally, in a scenario where Criteria 1 is not fulfilled the transformation $U$ leads to another representation of the graph states as illustrated in Figure.~\ref{fig:anafin}, where the resulting states do not exhibit EPR-like correlations, as it is shown in $\Gamma_{\text{imperfect,AB}}$. Any bipartite graph state is thus equivalent to a set of pure states, a set of $\Gamma_{\Ocentered}$, and a set of $\Gamma_{\text{imperfect,AB}}$. 



Using a numerical and analytical analysis we have provided criteria upon which entanglement routing is feasible in a complex graph state. We have expended previous work to encompass the study of large networks. This allows the identification of complex structure under which Entanglement Routing can be performed, and gives insights to the shape needed for multipartite real-world quantum information process. When our conditions are not met, ideal Entanglement Routing, as defined in this work, cannot be achieved. The two providers are therefore limited to extracting an imperfect entangled state between the two users, which is not an EPR state.

Nevertheless, this imperfect state may still be sufficient for the implementation of a quantum teleportation protocol. Indeed, such a state may exhibit two-way steering, thereby rendering it suitable for use in a quantum information protocol \cite{fan2022quantum}. If such a statement can be proven, then even graph states where only imperfect routing, as defined here, can be achieved, could be sufficient candidates for quantum communication protocols.

\section*{Code availability}
The code used to generate data presented in this paper is available to the interested reader in \cite{Fainsin_Entanglement_Routing_Simulation}.
\section*{Acknowledgments}
We sincerely thank M. Carlos Ernesto Lopetegui-González for insightful discussions.
This   work   was   supported   by   the   European   Research Council under the Consolidator Grant COQCOoN (Grant No.  820079).

\section*{Author Contributions}
The project was conceived and overseen from start to finish by VP. DF and AD worked hand in hand on the code construction and analysis. Both extracted from the data the new results presented in this paper. DF also worked on the symplectic eigenvalue distribution analysis. IK worked on the simple case of routing impossibility. MW provided substantial new theoretical insights during most of the project.

\section*{Additional information}

\begin{FlushLeft}
\textbf{Supplementary Information} can be found in Appendix \\

\textbf{Competing interests}: All authors declare no financial or non-financial competing interests. 
\end{FlushLeft}

\appendix 
\section{Gell-mann parametrization of unitaries}\label{sec:Gellmann}
At each generation the algorithm needs to generate several elements of $U(n)$ to evaluate their score with the fitness function. We achieve this using Gell-mann parametrization \cite{Bertlmann_2008}. Other parametrizations such as \cite{Zyczkowski_1994} or \cite{QR} were used but lacked computational efficiency for dense explorations of a large space of parameters. The main idea behind the Gell-Mann parametrization is to use exponentiation of traceless hermitian matrices to generate elements of $U(n)$. Indeed any unitary $U$ can be written
\begin{equation}
U=e^{iH}
\label{eq:R19}
\end{equation}
where $H$ is hermitian.
Then here we make use of the generalized Gell-Mann matrices as generators for the unitary group. The Gell-Mann basis is composed of $n^2-1$ matrices which are constructed as follows and comprehend three different categories. Through all this part $E_{i,j}$ is a matrix composed of zeros only except for a one in position $(i,j)$. It composed of three sets of symmetric $\lbrace\Lambda_{j,k}^s\rbrace_{1\leq j<k \leq n}$, antisymmetric $\lbrace\Lambda_{j,k}^a\rbrace_{1\leq j<k \leq n}$ and diagonal matrices $ \lbrace\Lambda_l\rbrace_{1\leq l \leq n-1} $ as follow 
\begin{equation}
\begin{aligned}
\Lambda_{j,k}^s & = E_{j,k}+E_{k,j} \\
\Lambda_{j,k}^a & = -i(E_{j,k}-E_{k,j}) \\
\Lambda_l & = \sqrt{\frac{2}{l(l+1)}}\left(\sum_{j=1}^{l}E_{j,j}-lE_{l+1,l+1}\right)
\end{aligned}
\label{eq:R20}
\end{equation}
Renaming all the matrices as $\Lambda_1,...\Lambda_{n^2-1}$ we can now construct any $n\times n$ hermitian matrix through this basis with our parameters $\lbrace\epsilon_i\rbrace_{i\in \{1,...n^2-1\}}$ and then obtain the corresponding unitary through exponentiation
\begin{equation}
    H = \sum_{i=1}^{n^2-1} \epsilon_i\Lambda_i
\label{eq:R21}
\end{equation}
and the coordinates $\lbrace\epsilon_i \rbrace_{i=1,..,n^2-1}$ of $H$ in the Gell-Mann basis are the parameters we optimize with the evolutionary algorithm.

\section{Covariance Matrix Adaptation Evolutionary Algorithm}\label{sec:AppA}
We give a detailed description of the so-called $(\mu,\lambda)$ Covariance Matrix Adaptation evolutionary algorithm where $\lambda$ is the number of generated mutants per generation, and $\mu$ are the individuals selected to become subsequent parents \cite{Roslund09}. $\lambda$ is designed to scale with the dimensionality $D$ of our problem such that : 
\begin{equation}
\begin{aligned}
	\lambda & = 4 + \ln(D) \\
	\mu & = \lfloor\frac{\lambda}{2} \rfloor
\end{aligned}
\label{eq:R5}
\end{equation}
To produce a new generation the algorithm first generates $\lambda$ perturbations $\vec{z_k}\sim \mathcal{N}(0,1)$ from the former control point $\langle\vec{x}\rangle^{(i)}$ combined with a multivariate normal distribution parametrized by the statistically learned covariance matrix $\mathbf{\Gamma}$ such as : 
\begin{equation}
\Phi_{\mathcal{N}}(\vec{z}) = \frac{1}{\sqrt{(2\pi)^{D} \det\mathbf{\Gamma}}} \exp\left(-\frac{1}{2}\vec{z}^{~T} \mathbf{\Gamma}^{-1} \vec{z}\right). 
\label{eq:R6}
\end{equation}
$\Gamma$ is a weighted history of the successful evolution path $\vec{p}_c$ in an effort to minimize the effects of noise. The evolution path is calculated this way :
\begin{equation}
\vec{p}_c^{(i+1)}=(1-c_c)\vec{p}_c^{(i)}+\sqrt{c_c(2-c_c)\mu_{eff}}\mathbf{R}^{(i)} \Lambda^{(i)1/2} \langle\vec{z}\rangle^{(i+1)},
\label{eq:R7}
\end{equation}
where $c_c = \frac{4}{D+4}$, $\mu_{eff} = \frac{1}{\sum_{k=1}^{\mu}\omega_k^2}$, $\mathbf{R}$ and $\Lambda$ come from the eigendecomposition of $\mathbf{\Gamma} = \mathbf{R}\Lambda \mathbf{R}^{t}$. The weights $\omega_k$ are defined in \eqref{eq:R15}. $\mathbf{\Gamma}$ is then updated according to:

\begin{multline}
\mathbf{\Gamma}^{(i+1)}= (1-c_{cov})\mathbf{\Gamma}^{(i)}+\frac{c_{cov}}{\mu_{eff}}\vec{p}_c^{(i+1)}\vec{p}_c^{(i+1)T} \\ +c_{cov}\left( 1-\frac{1}{\mu_{eff}} \right)\mathbf{R}^{(i)}\Lambda^{(i)1/2} \\ \times
\left(\sum_{k=1}^{\mu} \omega_k \vec{z}_k^{(i+1)} \vec{z}_k^{(i+1)T}\right)\left(\mathbf{R}^{(i)}\Lambda^{(i)1/2}\right)^{T}
\label{eq:R8}
\end{multline} 

Where $c_{cov}=\frac{2}{(D+\sqrt{2})^2}$ and $\mathbf{\Gamma}^{(0)}=\bm{1}$. This statistical covariance matrix is then coupled to a factor $\sigma_g$ to model an expansion or contraction of the main path. The idea is to begin by exploring a large diversity of offspring, to limit the influence of local minima and then progressively reducing the diversity to look for the precise solution of our problem, as illustrated in Figure~\ref{fig:CMA_algo}. It is updated as follow 
\begin{equation}
\sigma_g^{(i+1)} = \sigma_g^{(i)}\exp\left[ \frac{c_{\sigma}}{d_{\sigma}}\left(\frac{\norm{\vec{p}_{\sigma}^{(i+1)}}}{\langle\norm{\mathcal{N}(0,\mathbf{1})}\rangle}-1 \right) \right]
\label{eq:R9}
\end{equation}
with 
\begin{equation}
\begin{aligned}
c_{\sigma} & = \frac{(\mu_{eff}+2)}{(D+\mu_{eff}+3)}\\
d_{\sigma} & = 1+c_{\sigma}
\end{aligned}
\label{eq:R10}
\end{equation}
and 
\begin{equation}
p_{\sigma}^{(i+1)}=(1-c_{\sigma})\vec{p}_{\sigma}^{(i)}+\sqrt{c_{\sigma}(2-c_{\sigma}\mu_{eff})}\mathbf{R}^{(i)}\langle\vec{z}\rangle^{(i+1)}.
\label{eq:R11}
\end{equation}
Finaly the $\lambda$ trial solutions are 
\begin{equation}
\vec{x}_{k}^{(i+1)} = \langle\vec{x}\rangle^{(i)} + \sigma_g^{(i)} \mathbf{R}^{(i)}\Lambda^{(i) 1/2}\vec{z}_k^{(i+1)}
\label{eq:R12}
\end{equation}
thus we have our new generation of offsprings. We now only need to compute our future control point $\langle\vec{x}\rangle^{(i+1)}$. First we need to evaluate each offspring of the new generation to determine the $\mu$ best candidates via the objective function $f$. We reorder our offspring $(\vec{x}_1,...\vec{x}_{\lambda})$ such that : 
\begin{equation}
f(\vec{x}_1)<f(\vec{x}_2)<...<f(\vec{x}_{\lambda})
\label{eq:R13}
\end{equation}
then keeping only the $\mu$ best candidates we build 
\begin{equation}
\langle \vec{x}\rangle^{(i+1)} = \sum_{k=1}^{\mu} \omega_k\vec{x}_k^{(i+1)}
\label{eq:R14}
\end{equation}
where the weights are set to give a higher importance to offsprings with better fitness function
\begin{equation}
\omega_k=\frac{\ln\left[(\mu+1)/k\right]}{\sum_{j=1}^{\mu}\ln\left[(\mu+1)/j\right] }
\label{eq:R15}
\end{equation}

\section{Analysis of the square state}\label{sec:AppC}
\begin{figure}
\begin{center}
\includegraphics[scale=0.8]{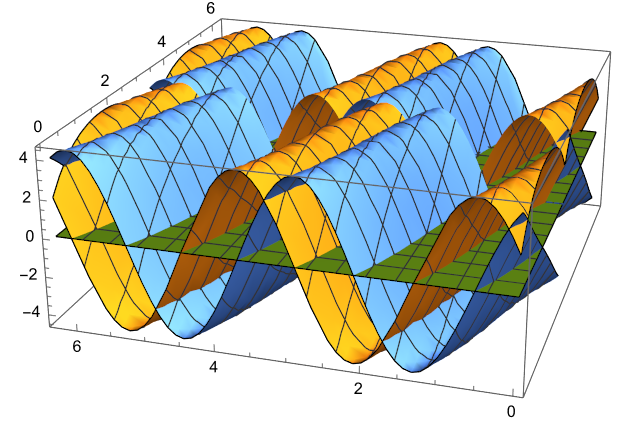}
\end{center}
	\caption{$\Delta^2\hat{X}_1 - \lambda$ et Cov($\hat{X}_1,\hat{P}_1$) pour $\phi_{1A} \in \left[0,2\pi\right]$ et $\phi_{2A} \in \left[0,2\pi\right]$. It appears that the two functions are never 0 at the same point of the space, thus showing the impossibility of a ideal routing scenario in square networks.}
	\label{fig:square_impos}
\end{figure}
The initial covariance matrix, \textbf{in the convention $(X_{A1},X_{A2},P_{A1},P_{A2},X_{B3},X_{B4},P_{B3},P_{B4})$}, is as follows

\begin{widetext}
\begin{equation*}
\Gamma_{\SmallSquarePoints} = 
\begin{pmatrix}
\frac{3s}{5}+\frac{2}{5s} & 0 & 0 & \frac{s^2-1}{5s} & 0 & -\frac{2(s^2-1)}{5s} & \frac{s^2-1}{5s} & 0\\
0 & \frac{3s}{5}+\frac{2}{5s} & \frac{s^2-1}{5s} & 0 & -\frac{2(s^2-1)}{5s} & 0 & 0 & \frac{s^2-1}{5s}\\
0 & \frac{s^2-1}{5s} & \frac{2s}{5}+\frac{3}{5s} & 0 & \frac{s^2-1}{5s} & 0 & 0 & \frac{2(s^2-1)}{5s}\\
\frac{s^2-1}{5s} & 0 & 0 & \frac{2s}{5}+\frac{3}{5s} & 0 & \frac{s^2-1}{5s} & \frac{2(s^2-1)}{5s} & 0\\
0 & -\frac{2(s^2-1)}{5s} & \frac{s^2-1}{5s} & 0 & \frac{3s}{5}+\frac{2}{5s} & 0 & 0 & \frac{s^2-1}{5s}\\
-\frac{2(s^2-1)}{5s} & 0 & 0 & \frac{s^2-1}{5s} & 0 & \frac{3s}{5}+\frac{2}{5s} & \frac{s^2-1}{5s} & 0\\
\frac{s^2-1}{5s} & 0 & 0 & \frac{2(s^2-1)}{5s} & 0 & \frac{s^2-1}{5s} & \frac{2s}{5}+\frac{3}{5s} & 0\\
0 & \frac{s^2-1}{5s} & \frac{2(s^2-1)}{5s} & 0 & \frac{s^2-1}{5s} & 0 & 0 & \frac{2s}{5}+\frac{3}{5s}
\end{pmatrix}.
\end{equation*}
\end{widetext}

One known method for correlating or decorrelating two nodes in the systems we are studying here is to use symmetric beam-splitters, so we set $\theta_A = \theta_B = \pi/4$ in the preliminary approach. Then, it is possible to find 'by hand' a set of values that satisfy the problem for all $(i, j)$ satisfying \eqref{eq:ana_pb}. Using this approach we are not able to fully recover the covariance matrix of an EPR pair, yet we find
\begin{equation}
\Gamma_{gen} = 
\begin{pmatrix}
\lambda_- & 0 & 0 & \mu\\
0 & \lambda_- & \mu & 0\\
0 & \mu & \lambda_+ & 0\\
\mu & 0 & 0 & \lambda_+
\end{pmatrix}
\label{eq:R_asym_lamb}
\end{equation}
with
\begin{equation}
\left\lbrace \mu = \frac{s^2 - 1}{\sqrt{5}s}, \lambda_\pm = \frac{(\sqrt{5}+5)s^2 \pm (\sqrt{5} + 5)}{10s} \right\rbrace
\end{equation}
It appears that the problem has no solution in the case of symmetric beam-splitters (with lambda fixed). This impossibility was also extended to a scenario where no assumption is made on $\theta_{A/B}$. Looking at the analytical expression of $\Sigma_{gen}$ we find 

\begin{equation}
    \begin{split}
            Cov(\hat{X}_1,\hat{X_2})  \propto \left[\sin(\theta_A)\cos(\theta_A)\sin(\phi_{1A})+\cos(2\theta_A)\right] \\ \times \sin(\phi_{1A}+\phi_{2A}+\phi_{3A}) \\
  \propto \left[\sin(\theta_A)\cos(\theta_A)\sin(\phi_{1A})+\cos(2\theta_A)\right] \\ \times \cos(\phi_{1A}+\phi_{2A}+\phi_{3A})
    \end{split}
\end{equation}

which leads to the condition 
\begin{equation}
\sin(\theta_A) \cos(\theta_A) \sin(\phi_{1A})+\cos(2\theta_A)=0
\end{equation}
and thus
\begin{equation}
\theta_A = \frac{1}{2} \cot ^{-1}\left(-\frac{\sin (\phi_{1A})}{2}\right)
\end{equation}
We can now analytically assess $\Delta^2 \hat{X}_1 -\lambda$ and $Cov(\hat{X}_1,\hat{P}_1)$, are two functions of $\phi_{1A}$ and $\phi_{2A}$ for the determined $\theta_A$. We plot these two functions in Fig.~\ref{fig:square_impos}. There are no values of $(\phi_{1A},\phi_{2A})$ for which both covariances vanish at the same point, indicating that the problem has no solution even in the case of an unbalanced beam-splitter. Overall, it appears that in the case of a square cluster, if our evolutionnary strategy does not converge perfectly it is indeed due to an impossibility to find a solution. However, this result cannot be generalized yet to all cluster topologies with $n>4$.

\bibliography{sn-bibliography}

\end{document}